\documentstyle[preprint,aps,axodraw]{revtex}
\tightenlines
\begin{document}

\draft
\title{Electromagnetic and gravitational decay of the Higgs boson}
\author{R.~Delbourgo\cite{Author1} and Dongsheng Liu\cite{Author2}}
\address{School of Mathematics and Physics, University of Tasmania,\\
 Hobart, Australia 7001}
\date{\today }
\maketitle

\begin{abstract}
The decays of a scalar particle, of either parity, into two photons or into
two gravitons are evaluated. The effective interactions are of the form 
$\phi FF,~\phi F\tilde{F}$ or $\phi RR,~\phi R\tilde{R}$; in particular, 
Higgs boson decay into gravitons cannot be adduced to an interaction 
$\phi R$, as has been recently claimed.
\end{abstract}


\narrowtext


In a very recent paper, Srivastava and Widom \cite{SW} have claimed that 
the decay width of the Higgs meson into two gravitons is given by 
$\sqrt{2}G_Fm_H^3/16\pi$. In their result, which they say stems from an 
effective interaction $\phi R/\langle\phi\rangle$, Newton's constant 
disappears and gets replaced by the Fermi constant, leading to a large 
magnitude for the process. If their result were true, it would be 
counterintuitive to the notion that gravitational interactions are 
miniscule in particle physics and it would imply that the Higgs mesons 
disappears {\em very quickly} into a puff of gravitational plus 
electromagnetric radiation! In this paper we show that their result is not 
right and we derive the correct magnitudes for the decay amplitudes, be 
the decaying particle scalar or pseudoscalar; our faith in the weakness 
of induced gravitational (and electromagnetic) effects is happily restored.

We consider parity-conserving decays of a $0^+$ or $0^-$ particle into two
photons ($\gamma$) or into two gravitons ($h$). A simple helicity amplitude 
analysis \cite{f1} shows that either process is governed by just one reduced
helicity amplitude, $\langle k,\lambda; k',\lambda'|S|0\rangle$, where 
$\lambda=\lambda'=1$ or 2, because of parity conservation. The same 
conclusion is reached by a covariant amplitude analysis, using the twin 
principles of gauge and general covariance. If $\varepsilon$ represents the 
external wavefunction of an incoming massless particle ($\gamma$ or $h$), 
we may write down the {\em unique} couplings,

\noindent
\underline{Scalar $\rightarrow \gamma(k,\lambda)\gamma(k',\lambda')$}
\begin{equation}
{\cal L}_{\phi\gamma\gamma}=g\varepsilon^{*\mu}_\lambda(k)
        \varepsilon^{*\rho}_{\lambda'}(k')T_{\mu\rho}(k,k');\quad
    T_{\mu\rho}(k,k')\equiv k'_\mu k_\rho - \eta_{\mu\rho}k\cdot k',
\end{equation}
\underline{Pseudoscalar $\rightarrow \gamma(k,\lambda)\gamma(k',\lambda')$}
\begin{equation}
{\cal L}_{\phi_5\gamma\gamma}=g_5\varepsilon^{*\mu}_\lambda(k)
\varepsilon^{*\rho}_{\lambda'}(k')k^\alpha k'^\beta
     \epsilon_{\mu\rho\alpha\beta},
\end{equation}
\underline{Scalar $\rightarrow h(k,\lambda)h(k',\lambda')$}
\begin{equation}
{\cal L}_{\phi hh}=G\varepsilon^{*\mu\nu}_\lambda(k)
        \varepsilon^{*\rho\sigma}_{\lambda'}(k')[T_{\mu\rho}(k,k')
  T_{\nu\sigma}(k,k')+T_{\nu\rho}(k,k')T_{\mu\sigma}(k,k')]/2;
\end{equation}
\underline{Pseudoscalar $\rightarrow h(k,\lambda)h(k',\lambda')$}
\begin{equation}
{\cal L}_{\phi_5 hh}=G_5\varepsilon^{*\mu\nu}_\lambda(k)
        \varepsilon^{*\rho\sigma}_{\lambda'}(k')
   [\epsilon_{\mu\rho\alpha\beta}k^\alpha k'^\beta T_{\nu\sigma}(k,k')
    + ({\rm 3~other~perms})]/4;
\end{equation}
Above, $\eta$ stands for the Minkowski metric, about which we have expanded 
the gravitational metric field ($g^{\mu\nu}=\eta^{\mu\nu}+\kappa h^{\mu\nu}$)
and with respect to which Lorentz scalar contractions are made.

There are four significant points about eqs (1) - (4). Firstly, we note that
insistence on current and stress tensor conservation means that each photon
is accompanied by one power of momentum and each graviton comes with two 
powers of momentum, entirely in keeping with low-energy theorems \cite{W}. 
Secondly, we can encapsulate the gauge invariances by regarding the 
effective couplings above as corresponding to effective gauge interactions,
\begin{equation}
 g\phi F^{\mu\nu}F_{\mu\nu}/4, \quad
 g_5\phi_5\epsilon_{\mu\nu\rho\sigma}F^{\mu\nu}F^{\rho\sigma}/8, \quad
 G\phi R_{\mu\nu\rho\sigma}R^{\mu\nu\rho\sigma}/2\kappa^2,\quad
 G_5\phi_5\epsilon_{\mu\nu\rho\sigma}{R^{\mu\nu}}_{\alpha\beta}
                                      R^{\rho\sigma\alpha\beta}/4\kappa^2,
\end{equation}
where $\kappa^2 = 8\pi G_{\rm Newton}$.
Thirdly, the mass dimensions of $g,g_5$ are $M^{-1}$ and of $G, G_5$ are 
$M^{-3}$. Fourthly and most importantly, the curvature tensor $R$ has only 
two powers of derivatives, but of course contains all powers of the graviton
field $h$, as befits a nonlinear theory. Thus it is quite impossible for an 
interaction like $\phi R$ to lead to two-graviton decay, as it is essential 
to conjure up {\em four} powers of momentum in the decay amplitude \cite{f2}.
In this respect we find the Srivastava-Widom claim extremely puzzling, 
notwithstanding that one is dealing with a spontaneously broken gauge 
symmetry and a vacuum expectation value which is closely tied to
the Higgs boson field.

It is straightforward to compute the decay rates ensuing from the 
interactions (1) to (4). Because gravitons and photons are massless the 
results are really quite simple in the end; if $m$ denotes the mass of the 
decaying scalar, one finds the partial widths,
\begin{equation}
\Gamma_{\phi\gamma\gamma}\!=\!|g|^2m^3\!/64\pi,\,\,
\Gamma_{\phi_5\gamma\gamma}\!=\!|g_5|^2m^3\!/64\pi,\,\,
\Gamma_{\phi hh}\!=\!3|G|^2m^7\!/512\pi,\,\,
\Gamma_{\phi_5 hh}\!=\! |G_5|^2m^7\!/512\pi.
\end{equation}
The first and third of these are relevant to the Higgs boson. The only 
remaining question is the magnitude of the various coupling constants, $g$ 
to $G_5$.

Because the decays are induced by quantum effects, they first arise at 
one-loop level. We may determine the couplings by applying the standard 
Feynman rules to the graphs depicted in Figure 1. In quoting the answers 
below, for a source field $\psi$ of mass $m_\psi$ circulating round the loop
which is either scalar or spinor and of unit charge $e$, we should point out
that certain graphs of Figure 1 are absent or give vanishing contributions; 
for example there is no Figure 1d for photons, but it certainly arises for 
gravitons; similarly Figure 1c does not occur for fermions undergoing 
electromagnetic decay but is present for gravitons and always exists
for boson sources. As a purely technical point, we have applied dimensional
regularization when calculating these diagrams, in order to respect gauge 
invariance.
\begin{mathletters}
\begin{equation}
g = \frac{e^2g_{\phi\psi\psi}m_\psi}{2\pi^2}\int_0^1\int_0^1
     dxdy\,\theta(1-x-y)\,\,(4xy-1)/(m_\psi^2-m^2xy);\quad {\rm fermion~loop}
\end{equation}
\begin{equation}
g = \frac{e^2g_{\phi\psi\psi}}{2\pi^2} \int_0^1\int_0^1
     dxdy\,\theta(1-x-y)\,\,xy/(m_\psi^2-m^2xy);\quad {\rm boson~loop}
\end{equation}
\end{mathletters}
\begin{equation}
g_5 = \frac{e^2g_{\phi_5\psi\psi}m_\psi}{2\pi^2}\int_0^1\int_0^1
     dxdy\,\theta(1-x-y)\,/(m_\psi^2-m^2xy); \quad {\rm fermion~loop~only}
\end{equation}
\begin{mathletters}
\begin{equation}
G = \frac{\kappa^2g_{\phi\psi\psi}m_\psi}{2\pi^2}\int_0^1\int_0^1
     dxdy\,\theta(1-x-y)\,\,xy(1-4xy)/(m_\psi^2-m^2xy);
    \quad {\rm fermion~loop}
\end{equation}
\begin{equation}
G = -\frac{\kappa^2g_{\phi\psi\psi}}{2\pi^2} \int_0^1\int_0^1
     dxdy\,\theta(1-x-y)\,\,x^2y^2/(m_\psi^2-m^2xy);\quad {\rm boson~loop}
\end{equation}
\end{mathletters}
\begin{equation}
G_5 = -\frac{\kappa^2g_{\phi\psi\psi}m_\psi}{2\pi^2} \int_0^1\int_0^1
     dxdy\,\theta(1-x-y)\,\,xy/(m_\psi^2-m^2xy);\quad {\rm fermion~loop~only}.
\end{equation}
Note that $g_{\phi\psi\psi}$ is dimensionless when $\psi$ is a spinor, but 
acquires dimensions of mass when $\psi$ is a scalar field. Anyhow, we see 
that all the integrals can be written as combinations of the basic integral,
\begin{equation}
 I_n(\mu) \equiv \int_0^1\int_0^1 dxdy\,\theta(1-x-y)\,(xy)^n/(1-\mu xy);
 \qquad \mu=m^2/m^2_\psi,
\end{equation}
which can be expressed as sums of dilog functions and the like \cite{f3}. As there
is little to be gained by looking up tables of Spence functions, it turns out
to be much easier to evaluate the $I_n$ numerically for input values of $\mu$.

Focussing on the Higgs boson now, and in respect of spin 0 and spin 1/2
contributions to the decay widths $\Gamma_{H\gamma\gamma}$ and 
$\Gamma_{Hgg}$, we should remind ourselves that $g_{H\psi\psi}=-2m^2_\psi/v$ for scalar
isosinglets and $g_{H\psi\psi} = -m_\psi/v$ for quarks, where $v = 246$ GeV 
is the Higgs vacuum expectation value. Therefore the couplings entering the 
partial widths (6) equal $g=-e^2I_1(\mu)/\pi^2 v$ and 
$G=-\kappa^2I_2(\mu)/\pi^2 v$ from a boson loop as well as
$g=e^2[I_0(\mu)-4I_1(\mu)]/\pi^2 v$ and 
$G=-\kappa^2[I_1(\mu)-4I_2(\mu)]/\pi^2 v$
from a fermion loop; in particular, a massless loop particle 
($\mu \rightarrow \infty$) reassuringly gives zero! Since the Higgs boson in
the standard model couples just to the quarks/leptons and gauge bosons, 
apart from its self-interactions, we get the fermionic contributions,
\begin{equation}
g\!=\!\frac{e^2}{2\pi^2 v}\!\left[\sum_{q=u,c,t}\!\!\frac{4}{3}\!
\left(I_0(m_q)\!-\!4I_1(m_q)\right)+\!\!\!\!\sum_{q=d,s,b}\!\!\frac{1}{3}\!
\left(I_0(m_q)\!-\!4I_1(m_q)\right)+\!\!\!\!\sum_{l=e,\mu,\tau}\!\!\!
\left(I_0(m_l)\!-\!4I_1(m_l)\right)\right]
\end{equation}
\begin{equation}
\!\!G\!=\!\frac{-\kappa^2}{2\pi^2 v}\!
\left[\sum_{q=u,c,t}\!\!\frac{4}{3}\!
\left(I_1(m_q)\!-\!4I_2(m_q)\right)+\!\!\!\!\sum_{q=d,s,b}\!\!\frac{1}{3}\!
\left(I_1(m_q)\!-\!4I_2(m_q)\right)+\!\!\!\!\sum_{l=e,\mu,\tau}\!\!\!
 \left(I_1(m_l)\!-\!4I_2(m_l)\right)\right]
\end{equation}
To complete the calculation for electroweak theory one should add the 
contributions from the W and Z gauge bosons, including ghost field 
contributions as required by the gauge-fixing scheme in use. These have been
computed previously \cite{DGK}, and are a factor of 21/4 greater than the
fermionic terms. But in any event we see that the gravitational decay rate 
is largely determined by the top quark (and the heavy gauge bosons), with
a magnitude $G \sim G_{\rm Newton}m_t^2/vm_H^2$ that is fixed by the 
Newtonian constant. This enters eq. (6) and duly gives a miniscule result 
for the gravitational decay rate, of the order of
$$\Gamma_{Hhh}\sim m_H \,(G_{\rm Newton}m_H^3/v)^2/500\sim 
m_H\times 10^{-75}.$$
This is as it should be: the disappearance of the Higgs boson into 
gravitational radiation will never be observed.

\acknowledgments
We wish to thank the Australian Research Council for providing financial 
support in the form of a large grant, \#A69800907 and have benefitted
from discussions with Peter Jarvis.

\newpage
\begin{center} {\bf FIGURE CAPTION} \end{center}

\noindent Figure 1. One-loop contributions to Higgs decay into two
massless gauge particles where the circulating loop particle can be scalar,
spinor or vector.

\begin{figure}
\pagestyle{empty}
\begin{center}\begin{picture}(600,200)(0,200)
\DashLine(20,146)(50,146){5}
\ArrowLine(50,146)(100,176)
\ArrowLine(100,116)(50,146)
\ArrowLine(100,176)(100,116) 
\Photon(100,176)(150,176){4}{4.5}
\Text(145,190)[]{$k$}
\Photon(100,116)(150,116){4}{4.5}
\Text(145,100)[]{$k'$}

\Text(74,90)[]{\bf (a)}
\Text(174,146)[]{+}
\DashLine(200,146)(230,146){5} 
\ArrowLine(230,146)(280,176)
\ArrowLine(280,116)(230,146)         
\ArrowLine(280,176)(280,116) 
\Photon(280,176)(340,116){4}{7.0}
\Text(335,190)[]{$k$}
\Photon(280,116)(340,176){4}{7.0}
\Text(335,100)[]{$k'$}

\Text(254,90)[]{\bf (b)}
\Text(0,6)[]{+}
\DashLine(20,6)(50,6){5}
\ArrowArcn(75,6)(25,0,180)
\ArrowArcn(75,6)(25,180,0)
\Vertex(100,6){1.5}
\Photon(100,6)(150,36){4}{4.5}
\Text(145,50)[]{$k$}
\Photon(100,6)(150,-24){4}{4.5}
\Text(145,-38)[]{$k'$}

\Text(74,-50)[]{\bf (c)}
\Text(174,6)[]{+}
\DashLine(200,6)(230,6){5}
\ArrowArcn(255,6)(25,0,180)
\ArrowArcn(255,6)(25,180,0)
\Photon(280,6)(300,6){4}{2.5}
\Vertex(300,6){1.5}

\Photon(300,6)(330,36){4}{4.5}
\Text(325,50)[]{$k$}
\Photon(300,6)(330,-24){4}{4.5}
\Text(325,-38)[]{$k'$}

\Text(254,-50)[]{\bf (d)}

\end{picture}
\end{center}
\vspace{5in}
\caption{R Delbourgo}
\end{figure}


\begin{references}
\bibitem{f3} The analytic results read $I_0=[{\rm dilog}{z_+}+
 {\rm dilog}{z_-}]/z$, $I_1 = [{\rm dilog}{z_+}+{\rm dilog}{z_-}-z/2]/z^2$,
 $I_2 = [{\rm dilog}{z_+}+{\rm dilog}{z_-}-z/2-z^2/24]/z^3$, where 
 $z_\pm \equiv 2/[1 \pm \sqrt{1-4/\mu}]$.
 Also worth noting are the limits $I_n(\mu=0)=(n!)^2/(2n+2)!$, 
 corresponding to a superheavy massive loop particle; and for $\mu\gg 1$,
$I_0(\mu) \simeq \log^2(\mu)/2\mu, I_1(\mu) \simeq -1/2\mu, 
 I_2(\mu) \simeq -1/6\mu$, corresponding to a light fermion loop.
\end{references}
\end{document}